# Generative Refinement: A New Paradigm for Determining Single Crystal Structures Directly from HKL Data

**Wenlin Luo**

State Key Laboratory of Coordination Chemistry, School of Chemistry and Chemical Engineering, Nanjing University, Nanjing 210023, P. R. China
GitHub: https://github.com/1701NJU
Email: luowenlin862@gmail.com

**Yi Yuan**

State Key Laboratory of Coordination Chemistry, School of Chemistry and Chemical Engineering, Nanjing University, Nanjing 210023, P. R. China
Email: fanrong066@gmail.com

**Chenghui Li**[*]

State Key Laboratory of Coordination Chemistry, School of Chemistry and Chemical Engineering, Nanjing University, Nanjing 210023, P. R. China
Email: chli@nju.edu.cn

**Yue Zhao**[*]

State Key Laboratory of Coordination Chemistry, School of Chemistry and Chemical Engineering, Nanjing University, Nanjing 210023, P. R. China
Email: zhaoyue@nju.edu.cn

**Jing-Lin Zuo**[*]

State Key Laboratory of Coordination Chemistry, School of Chemistry and Chemical Engineering, Nanjing University, Nanjing 210023, P. R. China
Email: zuojl@nju.edu.cn

December 3, 2025

**Abstract**

Single-crystal X-ray diffraction (SC-XRD) is the "gold standard" technique to characterize crystal structures in solid state. Despite significant advances in automation for structure solution, the refinement stage still depends heavily on expert intervention and subjective judgment, limiting accessibility and scalability. Herein, we introduce *RefrActor*, an end-to-end deep learning framework that enables crystal structure determination directly from HKL data. By coupling a physics-informed reciprocal-space encoder (*ReciEncoder*) with a symmetry-aware diffusion-based generator (*StruDiffuser*), *RefrActor* produces fully refined atomic models without requiring initial structural guesses or manual input. Comprehensive evaluations on the **GenRef-10k** benchmark demonstrates that *RefrActor* achieves low $R_1$-factors across diverse systems, including low-symmetry, light-atom, and heavy-atom crystals. Case studies further confirm that *RefrActor* can correctly resolve hydrogen positions, elemental assignments, and moderate disorder. This work establishes a new data-driven paradigm for autonomous crystallographic analysis, offering a foundation for fully automated, high-throughput crystal structure determination.

## 1 Introduction

Determining the three-dimensional arrangement of atoms in crystalline solids lies at the heart of molecular science, as atomic configuration dictates nearly all physical, chemical, and biological behaviors of matter[1-7]. For over a century, single-crystal X-ray diffraction (SC-XRD) has remained the gold standard for this purpose, offering unparalleled insights that have shaped modern science[8-11]. Its impact is profound, having supported by Nobel Prize–winning work across multiple domains. For instance, Dorothy Crowfoot Hodgkin's award in 1964 for determination of the structures of penicillin and vitamin $B_{12}$ [12], and William N. Lipscomb's award in 1976 for revealing the bonding nature of boron hydrides[13]. Today, beyond these landmark achievements, SC-XRD is an indispensable tool in routine research, enabling the verification of reaction mechanisms, the rational design of pharmaceuticals, and the atomic-level engineering of advanced materials with tailored electronic, magnetic, or porous properties[14-20]. Despite its central role, however, one pivotal step in the SC-XRD workflow—structure refinement—remains a persistent bottleneck, slowing the pace of discovery and limiting accessibility for non-experts[21-25].

The workflow for crystal structure determination can be broadly divided into two principal stages: structure solution and structure refinement. Over the past few decades, structure solution has become highly automated, driven by advances in direct methods, Patterson techniques, and dual-space algorithms that efficiently generate initial atomic models from diffraction intensities[26-30]. In contrast, structure refinement remains a major bottleneck. This stage typically involves a semi-manual, iterative optimization process in which the crystallographer carefully adjusts atomic positions, thermal parameters, and occupancies using specialized programs such as *SHELXT*[31], *SHELXL*[32], or *OLEX2*[33]. Addressing issues such as atom misidentification, disorder, partial occupancy and anomalous scattering often demands expert interpretation and manual correction—tasks that remain inaccessible to most non-specialists. The reliance on expert-driven correction contrasts sharply with the trend toward high-throughput and automated research, highlighting a growing disparity between our ability to rapidly synthesize new compounds and our capacity to efficiently characterize their crystal structures.

Previous efforts toward automated crystal structure determination have largely centered on Crystal Structure Prediction (CSP), which formulates the problem as a computational search for low-energy atomic configurations[34-37]. Although effective in principle, such geometry-optimization–based approaches are computationally prohibitive for large or complex systems. Recent progress driven by machine learning and diffusion-based generative models—notably CDVAE and DiffCSP—has markedly improved sampling efficiency and structural fidelity[38-42]. However, these approaches share a fundamental limitation: they are trained to generate thermodynamically favorable structures solely from chemical composition, without leveraging the high-resolution reciprocal-space information embedded in experimental diffraction data. Consequently, they cannot directly refine candidate structures to reproduce observed diffraction patterns, and frequently fail to capture the experimentally realized polymorphs stabilized under specific crystallization conditions. The substantial challenge—to automatically reconstruct and refine atomic structures directly from X-ray diffraction data—thus remains unresolved.

Herein, we reconceptualize crystal structure refinement as a generative modeling problem and introducing a new paradigm we term **Generative Refinement**. To realize this concept, we developed ***RefrActor***, an end-to-end deep learning framework capable of directly producing fully refined crystal structures from experimental diffraction data, without requiring initial models or human intervention. *RefrActor* generates chemically consistent and diffraction-constrained atomic models that achieve high fidelity to both the measured intensities and the underlying bonding principles, enabling reliable reconstruction of structurally complex or disordered systems. By coupling a physics-informed diffraction encoder with a symmetry-aware generative architecture, *RefrActor* bridges the longstanding divide between automated structure solution and publication-ready refinement. This work constitutes the first demonstration of generative refinement for large-scale, automated interpretation of single-crystal diffraction data, opening new avenues toward autonomous crystallographic analysis and independent scientific discovery.

## 2 Result and discussion

### 2.1 The Generative Refinement Paradigm

In typical single crystal structure determination process, a crystal is mounted on a goniometer and illuminated by a collimated monochromatic X-ray beam, and the positions and intensities of diffracted beams are measured. A structural model can be generated based on the scattering amplitude and phase information. This model is then iteratively refined to minimize the discrepancy between the calculated and experimental diffraction data, thereby improving the agreement between the observed and simulated structure factors. The relationship between the three-dimensional atomic coordinates and the measured diffraction pattern is strictly governed by the Fourier transform of the electron density, forming a highly complex, nonlinear mathematical mapping. As a result, it is highly challenging for non-specialist users and automated computational programs to directly obtain precise crystal structure from the X-ray diffraction data.

Fortunately, the advancements in deep learning have allowed us to solve various problems in a "black-box" but autonomous manner. Recently, Madsen et al developed a neural framework to address the long-standing phase problem[43]. By directly taking the HKL indices and corresponding structure factor amplitudes as inputs, PhAI can infer the missing phase information and rapidly reconstruct initial electron density maps—circumventing the need for traditional, mathematically intensive phase-retrieval procedures. Despite this significant advance, PhAI is confined to the phase determination stage and lacks the capability for full structure refinement, a process that still demands expert supervision and iterative optimization. It is highly desirable to establish an end-to-end generative framework capable of producing fully refined crystal structures directly from experimental X-ray diffraction (HKL) data, thereby enabling non-expert users to obtain publication-ready atomic models without human intervention.

Based on this objective, the architecture of *RefrActor* is designed as an end-to-end framework that concretizes the paradigm of Generative Refinement. As illustrated in **Fig. 1**, it consists of two primary, interconnected modules: the ***ReciEncoder***, a physics-informed encoder that interprets reciprocal-space diffraction data, and the ***StruDiffuser***, a conditional generative model that reconstructs the corresponding real-space atomic configuration. This dual-module design establishes a continuous information pathway from raw experimental data to fully refined crystal structures. Detailed architectural specifications and training procedures are provided in the Supplementary Information (SI).

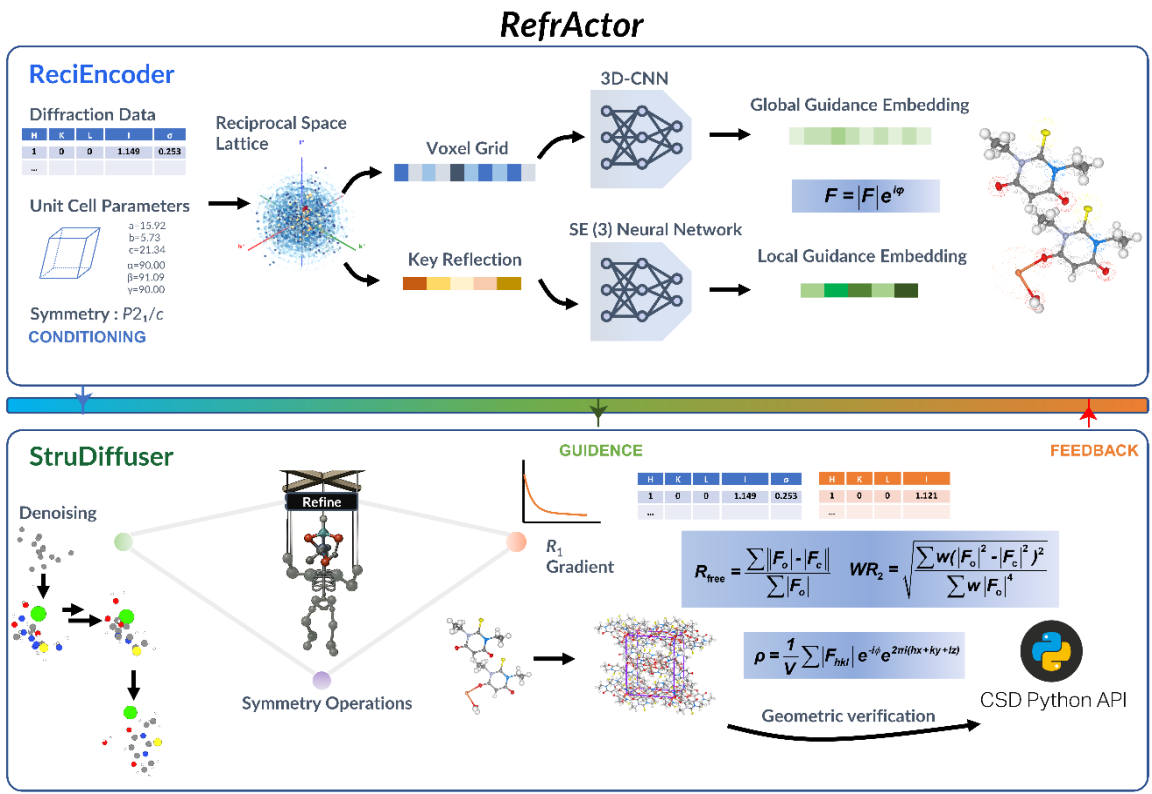

**Fig. 1 Overview of *RefrActor*.**

**2.2 Preparation of High-Fidelity Experimental Dataset**

The foundation of our generative refinement framework lies in a large-scale, high-fidelity dataset meticulously curated from established crystallographic databases, including the Cambridge Structural Database (CSD)[44] and Crystallography Open Database (COD)[45]. Our goal was to establish a robust correspondence between experimental diffraction measurements (HKL files) and their ground-truth crystal structures (CIFs). A cornerstone of this process is the rigorous crystallographic treatment of each structure. Using Pymatgen[46] and its *SpacegroupAnalyzer* module, we first reconstructed the fully symmetrized unit cell for each CIF. This critical step ensures that all symmetry-equivalent atomic sites were explicitly represented. Subsequently, a comprehensive bond network was constructed for every structure. Most importantly, we identify the atoms constituting the **asymmetric unit (ASU)** and establish a definitive mapping from every atom in the full cell back to its ASU representative. This pipeline yields a physically meaningful, symmetry-aware structural ground truth that serves as the foundational scaffold for our 'marionette' generative model, ensuring that all generated atomic configurations are crystallographically valid and chemically interpretable.

With a well-defined reference structure established, the next step was to transform the corresponding experimental HKL data into a standardized, physically informed representation suitable for generative learning. To achieve this, we employed the Computational Crystallography Toolbox (CCTBX)[47], a professional-grade suite for crystallographic data analysis. Each HKL file was parsed into a CCTBX *Miller array* object, a structured data container that intrinsically links the measured diffraction intensities with the crystal's unit cell parameters and symmetry operations. A critical aspect of our pipeline is the intelligent processing of these reflections. Rather than using raw intensities, we convert them to normalized structure factor amplitudes, accounting for the expected intensity decay with resolution (i.e., Wilson B-factor) on a per-sample basis. This ensures that the model learns from a physically meaningful representation of diffraction strength. To ensure equivariance under the Special Euclidean group SE(3), the encoding pathway implements a reflection selection strategy that identifies a subset of representative reflections for downstream learning. Beyond simple intensity-based filtering, reflections were prioritized using a composite scoring metric integrating their statistical significance ($F/\sigma$), resolution shell, and crystallographic relevance. This approach yields a sparse yet information-complete representation of the reciprocal space, maximizing both computational efficiency and structural fidelity.

The final stage of the data pipeline involves organizing the processed diffraction information into a representation that fully exploits the hybrid architecture and learning dynamics of *RefrActor*. To support the dual-path way, each sample's diffraction data is encoded into two complementary modalities: a 3D voxelized reciprocal lattice, optimized for the convolutional branch to capture global statistical features, and a point cloud of key reflections, enabling the SE(3) equivariant network to model precise, orientation-invariant geometric relationships. To enhance the discriminative capability of the encoder, we further incorporated a contrastive learning objective, constructing structurally similar positive pairs and dissimilar negative pairs for each sample. This objective encourages the encoder to learn an embedding space in which distinct diffraction fingerprints correspond uniquely to specific crystal structures. Collectively, this pipeline yields a multimodal and self-supervised dataset, where each sample integrates ground-truth structural information, dual diffraction representations, and contrastive supervision—providing the essential foundation for end-to-end training of the *RefrActor* framework.

## 2.3 *ReciEncoder* Performance on Experimental Diffraction Data

The efficacy of our generative refinement paradigm (**Fig. S1 and S2**) hinges on the quality of the guidance embedding produced by the *ReciEncoder*, which must encapsulate rich and physically meaningful crystallographic information directly from raw diffraction patterns. To this end, we designed a hybrid dual-pathway architecture (**Fig. 2a**) that integrates a global 3D convolutional network—responsible for extracting large-scale statistical regularities of reciprocal space—with a local SE(3)-equivariant branch that encodes fine-grained, orientation-invariant geometric correlations (**Fig. S3**, **Table S3–S4**). The encoder is optimized through a progressive multi-task curriculum that reflects the sequential reasoning of crystallographic analysis (**Fig. 2b**). This staged learning strategy enables the model to acquire hierarchical crystallographic priors, yielding a robust and physically grounded embedding space that effectively guides the downstream generative refinement.

The initial "foundation" phase of the curriculum employs contrastive learning to train the model to differentiate among crystal structures and establish a discriminative embedding space. The effectiveness of this phase is evidenced by multiple complementary metrics. The cross-modal similarity matrix (**Fig. 2d**) exhibits a sharply defined bright diagonal, reflecting a precise alignment between the HKL- and structure-derived embeddings (**Fig. S4**). Moreover, t-distributed Stochastic Neighbor Embedding (t-SNE) projections reveal a physically interpretable latent space, characterized by distinct clustering across crystal systems (**Fig. 2e**) and a continuous gradient corresponding to unit-cell anisotropy (**Fig. 2f**). Further analyses confirm systematic organization with respect to atomic multiplicity, cell volume, and reflection density (**Figs. S7–S9**), demonstrating that the encoder internalizes crystallographic constraints in a physically consistent manner. Importantly, the *ReciEncoder* also displays fine-grained discriminative resolution, as shown in a representative structure retrieval test (**Fig. 2g**; **Figs. S10–S11**; **Table S5**), where it assigns the highest similarity score (0.9814) to the correct structure while appropriately ranking a closely related, yet distinct, negative sample with a lower score (0.9091).

Following the foundational phase, the curriculum advances to increasingly complex tasks encompassing atomic arrangement and diffraction physics. The model's proficiency in capturing interatomic geometry is demonstrated by its accurate prediction of interatomic vector distributions, yielding a high correlation coefficient (0.933) and a low root mean square error (RMSE, 0.025 Å) for the radial Patterson function(**Fig. 2h; Fig. S12**). For the critical phase classification task, the *ReciEncoder* achieved an area under the curve (AUC) of 0.952 (**Fig. 2i; Figs. S13–S15**), confirming its strong capability to discern symmetry-related features such as systematic absences within diffraction data. To evaluate the contribution of each architectural component, we performed a systematic ablation analysis (**Fig. 2c**). The hybrid model consistently and significantly outperformed both single-pathway (CNN-only and SE(3)-only) variants across all benchmark tasks. This provides compelling evidence of the complementary nature of the two pathways: the convolutional neural network (CNN) branch effectively captures global statistical correlations in reciprocal space, while the SE(3)-equivariant branch encodes orientation-independent local geometric relationships. Their integration is thus essential for constructing a high-fidelity guidance embedding that enables successful Generative Refinement.

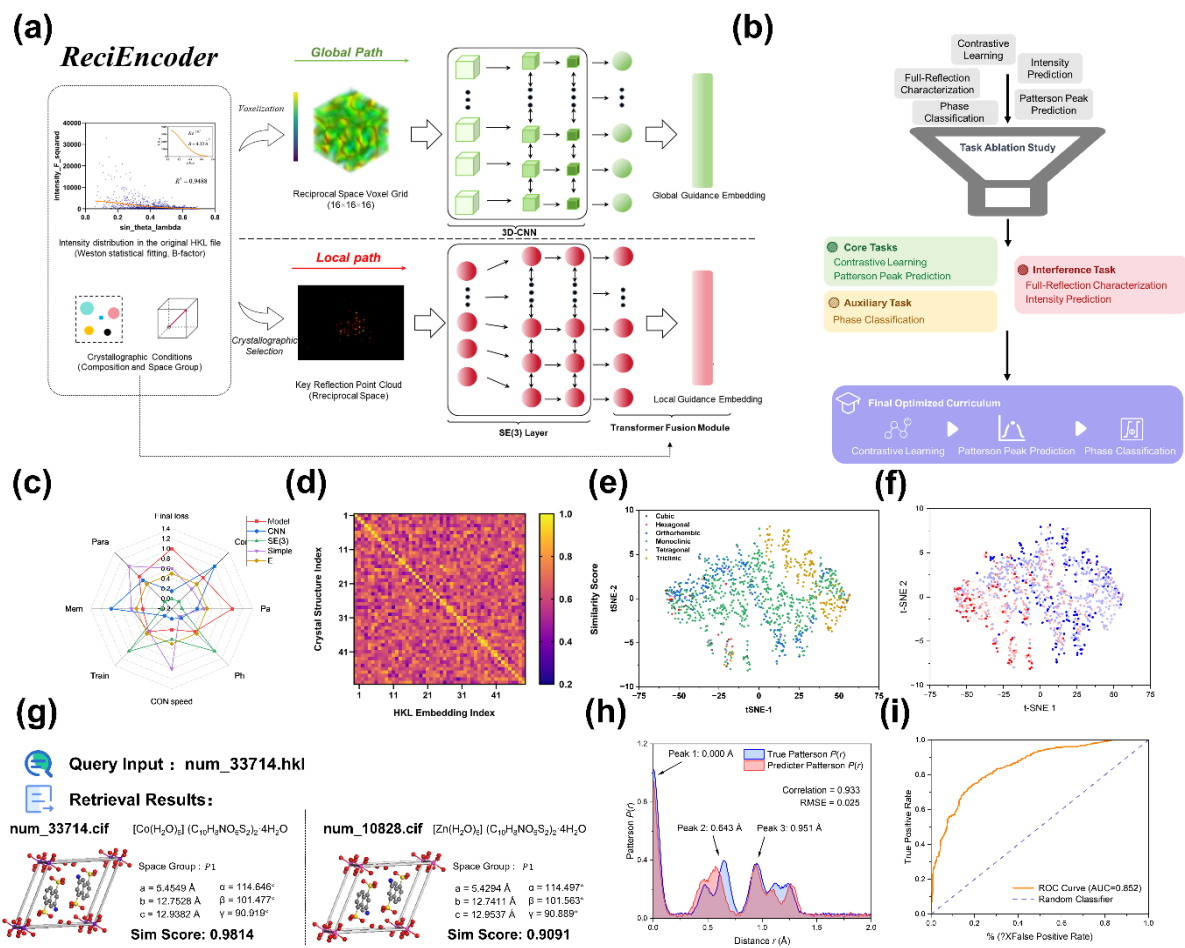

**Fig. 2** *ReciEncoder* **Architecture, Training, and Performance Evaluation. (a)** The hybrid architecture of the *ReciEncoder*, featuring a dual-pathway design composed of a Global Path (3D-CNN) and a Local Path (SE(3)-equivariant Transformer). **(b)** The optimization of the training strategy, illustrating how the final progressive multi-task curriculum was derived from a comprehensive task ablation study. **(c)** Radar chart comparing the performance and efficiency of different architectural variants. **(d)** Heatmap of the cross-modal similarity matrix from the contrastive learning task, showing strong diagonal alignment. **(e, f)** t-SNE visualizations of the learned embedding space, demonstrating physically meaningful organization based on (e) crystal system and (f) cell anisotropy. **(g)** Representative example of the retrieval task, showcasing the model's ability to identify the correct crystal structure (left) and a structurally similar one (right) from a query HKL, with corresponding high similarity scores. **(h)** Comparison of the true (blue) and predicted (red) radial Patterson function, indicating a high correlation (0.933) and low RMSE (0.025 Å). **(i)** The Receiver Operating Characteristic (ROC)

curve for the phase classification task (AUC = 0.852).

**2.4 Accuracy and Generalization Validation on GenRef-10k.**

Building on this modular foundation, the *StruDiffuser*—an enhanced generative backbone adapted from the SymmCD architecture[48]—coordinates with the *ReciEncoder* to enable fully end-to-end refinement. With this integrated architecture established, we proceeded to evaluate the holistic performance of the *RefrActor* framework. The central scientific question we sought to answer is whether our Generative Refinement paradigm can reliably and accurately produce high-quality crystal structures directly from experimental diffraction data across a vast and diverse chemical space. The definitive performance indicator is the crystallographic R-factor ($R_1$)—the standard metric quantifying the agreement between the generated structural model and the experimental diffraction intensities. A lower R-factor reflects a closer correspondence to experimental data, thereby signifying a refined structure that faithfully reproduces the true atomic arrangement.

To this end, we constructed **GenRef-10k**, a comprehensive benchmark dataset comprising approximately 10,000 experimentally derived crystal structures curated from multiple crystallographic databases, following the data processing pipeline shown in **Fig. 3a**. The GenRef-10k benchmark was deliberately constructed to encompass the full diversity of crystalline materials, spanning a wide range of chemical compositions (**Fig. 3b**), structural complexities (**Fig. 3c**), and symmetry classes (**Fig. 3d**). Supplementary analyses further confirm the dataset's compositional and crystallographic diversity, including balanced distributions of compound types, element counts, centrosymmetric and non-centrosymmetric structures, crystal densities, unit-cell and atomic volumes, and reflection statistics (**Figs. S16– S22, Table S6 and S7**). Collectively, these statistics establish GenRef-10k as a rigorous and representative benchmark for evaluating data-driven crystallographic models. On this benchmark, *RefrActor* demonstrated outstanding end-to-end performance. The distribution of the final $R_1$-factors, shown in **Fig. 3e**, reveals that a significant majority of the structures generated by *RefrActor* fall within the range typically considered well-refined ($R_1 < 5\%$). This result provides the primary evidence that our automated, end-to-end approach is not merely a theoretical construct, but a practical and powerful tool capable of consistently producing high-fidelity crystal structures.

To further evaluate the robustness and generalizability of our framework, we tested *RefrActor*'s on a series of challenging subsets within the GenRef-10k benchmark that are known to pose difficulties for conventional refinement algorithms. These subsets included crystals with low-symmetry space groups, structures composed exclusively of light elements, and those dominated by heavy atoms (**Table S8**). As summarized in **Fig. 3f** and further supported by the corresponding statistical distributions in **Fig. 3g**, *RefrActor* maintained consistently high refinement accuracy across all categories. The median $R_1$ factor remained low, with a large fraction of structures in each subset achieving $R_1 < 5\%$ (**Table S9**), underscoring the model's robust performance even in regimes that traditionally challenge automated refinement. This consistent behavior across diverse and demanding crystallographic conditions highlights the core advantage of our generative, data-driven paradigm. By learning the complex, non-linear relationships between diffraction intensities and atomic configurations from a broad and chemically diverse dataset, *RefrActor* can navigate refinement landscapes that often confound methods reliant on simpler heuristics or initialization-dependent models.

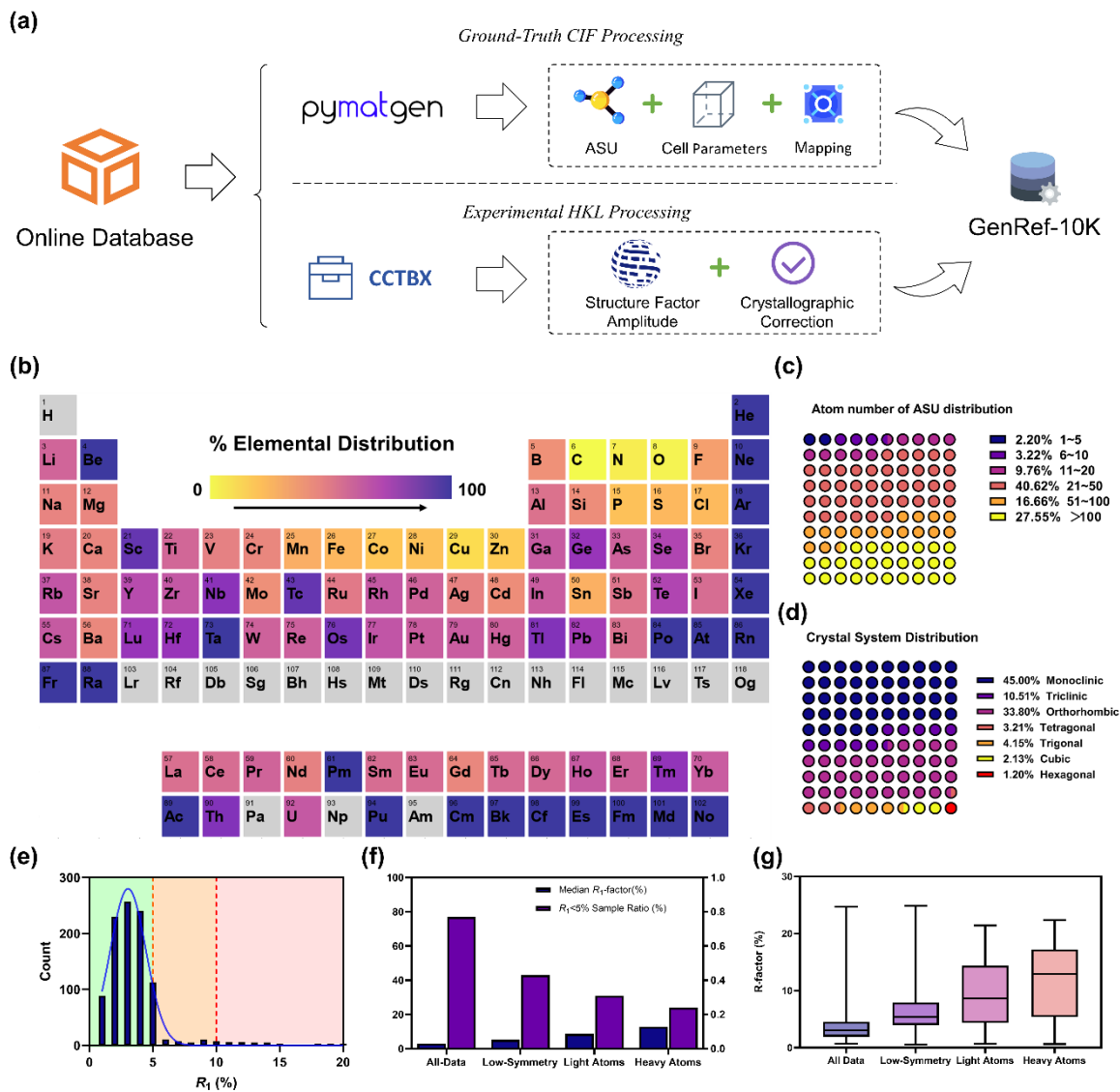

**Fig. 3 GenRef-10k Dataset Statistics and *StruDiffuser* Performance.** (a) The data processing pipeline for constructing the GenRef-10k benchmark, illustrating the dual-stream processing of ground-truth structures (via Pymatgen) and experimental diffraction intensities (via CCTBX) from online repositories. (b) Heatmap of elemental prevalence within GenRef-10k visualized on the periodic table, highlighting broad chemical coverage. (c, d) Statistical breakdown of the dataset by (c) atom count in the asymmetric unit (ASU) and (d) crystal system classification. (e) Histogram of the final crystallographic $R_1$-factors achieved on the test set, overlaid with a distribution curve. The green shaded region ($R_1 < 5\%$) denotes the high-quality refinement regime, where the majority of structures are located. (f) Comparative performance analysis across challenging subsets (Low-Symmetry, Light Atoms, and Heavy Atoms), reporting both the median $R_1$-factor and the success rate (ratio of samples with $R_1 < 5\%$). (g)

Box and whisker plots detailing the $R_1$-factor distributions for the full dataset and the challenging subsets, demonstrating robust performance stability across diverse crystallographic conditions.

**2.5 Case Study on Refining Single Crystal Structure**

In the following subsections, we evaluate the feasibility of the performance of *RefrActor* by testing several crystal structures which are not part of the GenRef-10k. To ensure comprehensive evaluation, we deliberately selected a range of crystallographic materials, encompassing organic molecules, inorganic clusters, and metal-organic complexes. The structures generated by *RefrActor* were compared with their corresponding crystallographer-refined reference structure to assess the performance metrics. For clarity, the structures refined by crystallographers, generated by our method, and their overlay are presented in the left, middle, and right columns, respectively (**Figs. 4a–d**). As a baseline demonstration of accuracy, the crystal structures generated by our method (e.g., **Figs. 4a, b**; **Figs. S25-27**) demonstrate that nearly all tested datasets can accurately reproduce the structural composition, even without prior knowledge of elemental constituents. Remarkably, the AI-derived structural units for these inorganic and organic examples correctly identified the elemental species, with negligible deviations from the crystallographer-refined structures. In certain cases, our method outperforms existing crystallographic refinement programs, including *SHELXT*, the most widely used and powerful structure solving tool. As demonstrated in **Fig. 4e**, *SHELXT*, despite being provided with the correct elemental composition, misassigned atomic species (e.g., O as C/N, or N as C). Such errors typically require post correction by crystallographers, based on chemical intuition and crystallographic expertise. By contrast, our approach directly generates the correct structural model, with atomic assignments fully consistent with the target compound. These results highlight a clear advantage over existing crystallographic software, suggesting that our method could reduce manual intervention and improve the reliability of structural determination.

As hydrogen atoms only scatter X-radiation weakly, it is difficult to precisely locate the position of hydrogen from X-ray diffraction. Therefore, hydrogen atoms are commonly add manually by crystallographers in SC-XRD. Notably, the AI-derived structures not only correctly identified all atomic species but also precisely determined the positions of hydrogen atoms. The method can also accurately predict the hydrogen-bonding networks formed between hydrogen atoms and their surrounding atoms. Most impressively, our method can precisely determines the stoichiometry of hydrogen atoms, a particularly challenging task given that the position and number of H-atoms in metal-organic complexes are often governed by the oxidation state of the metal center. In conventional crystallographic refinement, such determinations typically require manual intervention based on metal valence considerations. A striking example is the AI-generated structure of $PbBr_4C_{18}H_{28}N_2$ (**Fig. 4c**), which correctly assigns both the positions and quantities of H atoms. This system exemplifies the delicate charge balance in perovskite-like frameworks, where the anionic PbBr skeleton carries negative charge compensated by cationic organic moieties. The method's ability to accurately predict H-atom configurations in such systems suggests that it possesses emerging capability to infer the oxidation states and charge distribution in metal-organic compounds.

Moreover, *RefrActor* could generate the exact molecular structure for slightly disordered molecules. As show in **Fig. 4d**, in the AI-generated metal-organic framework (MOF) structures, the pyridyl groups in the ligands exhibit rotational disorder, a phenomenon traditionally resolved by crystallographers through iterative refinement of electron density maps. Remarkably, our method autonomously identifies the disordered configurations without manual intervention. As evidenced by the structural overlay (see **Fig. 4b**, right), the AI-predicted disordered positions show near-perfect alignment with those refined by crystallographers. The capability to accurately resolve ligand disorder, a common challenge in MOF crystallography, highlights a distinct advantage of our approach.

## (a) $Cs_3Cu_2I_5$

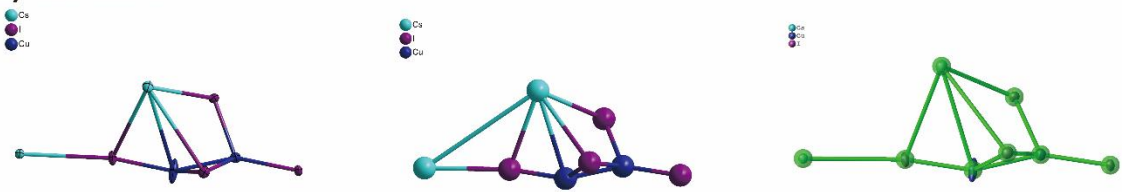

## (b) $C_{12}H_{16}N_6O_4$

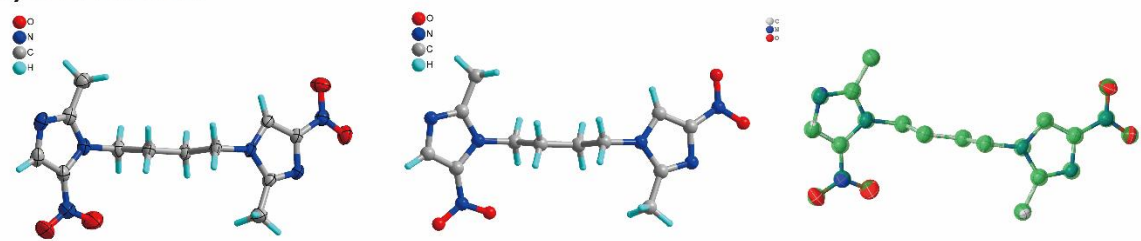

## (c) $PbBr_4C_{18}H_{28}N_2$

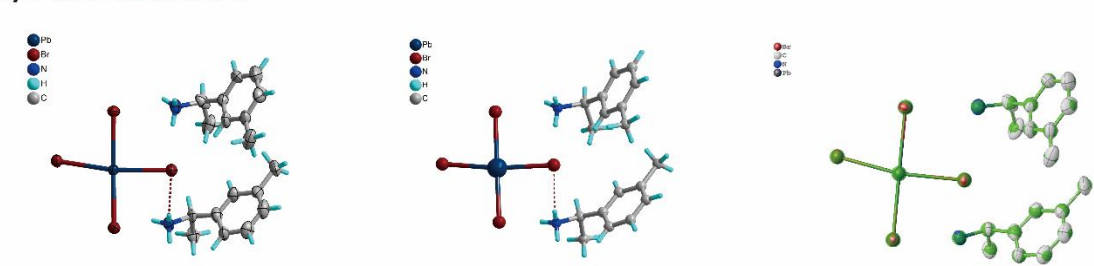

## (d) $C_{105}H_{75}N_9Ni_4O_{13}S$

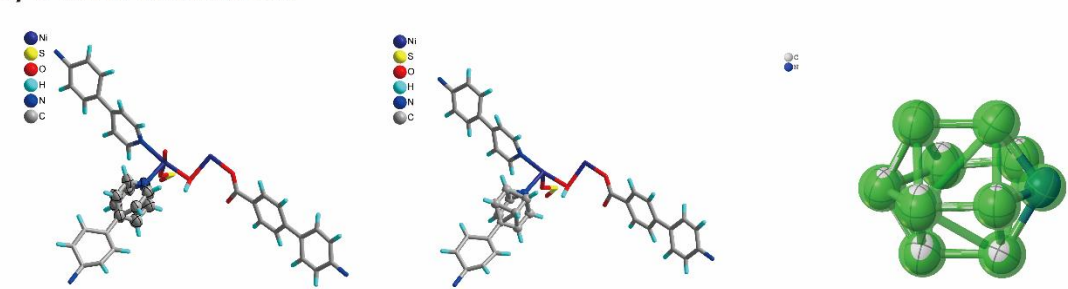

## (e) $C_7H_4N_{10}O_8$

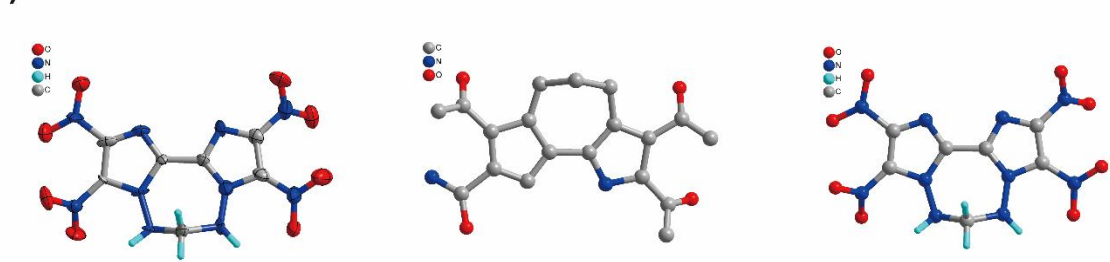

**Figure 4** Structural comparison of refinement outcomes. (a–d) Comparison between AI-generated and crystallographer-refined structures. Each panel follows a three-column layout showing the crystallographer-refined structure (left), the AI-generated structure (middle), and their overlay (right). (e) Comparison of refinement methods, where the left, middle, and right columns correspond to the crystallographer-refined, *SHELXT*-generated, and AI-generated structures, respectively. Case studies: **(a)** Inorganic system ($Cs_3Cu_2I_5$); **(b)** Organic system ($C_6H_{30}O_{10}$); **(c)** Metal–organic complex ($PbBr_4C_{18}H_{28}N_2$) highlighting H-atom placement; **(d)** Disordered system ($C_{105}H_{75}N_9Ni_4O_{13}S$); **(e)** Atomic assignment comparison with *SHELXT* ($C_7H_4N_{10}O_8$).

## 2.6 Intended Use and Limitations

We intend to use our method to deduce refined structures directly from the corresponding HKL data even in the absence of the unit cell and elemental compositions. This approach significantly simplifies the process, offering particular convenience for someone without specialized crystallographic expertise. The computational workflow comprises three key steps: (1) loading the HKL diffraction data, (2) inferring the space group using *ReciEncoder*, and (3) generating candidate structures via *StruDiffuser*. The resulting structures are ranked by their $R_1$ values, where a lower value typically indicates greater consistency between the predicted crystal structure and the experimental HKL data. Accordingly, the structure with the lowest value can be regarded as the refined structure. Consequently, *RefrActor* empowers researchers—regardless of their background in crystallography—to generate high-quality structures and CIF files directly from raw experimental data.

Despite the remarkable progress enabled by *RefrActor*, several intrinsic challenges remain at the interface between crystallography and deep learning. First, unambiguously identifying space groups from diffraction data remains non-trivial for the *ReciEncoder*. Distinct symmetries can yield nearly indistinguishable signatures in reciprocal space—particularly when data quality deteriorates—introducing upstream ambiguity that propagates downstream and limits structural fidelity. Second, the framework faces constraints when handling complex crystallographic phenomena such as twinning and severe disorder. These cases violate the model's current assumptions of single, ordered domains and reveal the absence of explicit mechanisms to deconvolute composite diffraction patterns or represent atomic positions probabilistically. Finally, for crystals containing heavy elements, reconstruction accuracy can be constrained by the physics of X-ray scattering: heavy atoms dominate the scattering intensity and may suppress contributions from lighter atoms. This imbalance has the potential to produce skewed feature gradients that bias the network, an issue further exacerbated by the statistical predominance of organic structures in the training corpus. Overcoming these challenges will require more expressive reciprocal-space encoders, generative paradigms capable of modeling multi-domain systems, and training strategies grounded in physical constraints. Together, these advances will move data-driven crystallography closer to a truly general framework for structural discovery.

## 3 Conclusion

In this work, we present *RefrActor*, a deep learning framework that establishes a new paradigm for crystal structure determination—generative refinement. By tightly integrating a physics-informed reciprocal-space encoder (*ReciEncoder*) with a symmetry-aware generative model (*StruDiffuser*), *RefrActor* reconstructs high-fidelity crystal structures directly from experimental diffraction data, effectively eliminating the need for initial models or prior chemical assumptions. Comprehensive validation on the large-scale GenRef-10k benchmark demonstrates that *RefrActor* achieves state-of-the-art accuracy and generalizability across a broad chemical and crystallographic landscape. Notably, the model retains robust performance under conditions that traditionally challenge refinement—such as low symmetry, light-atom systems, and heavy-atom dominance. Through detailed case studies, we further show that *RefrActor* can replicate expert-level results and, in certain cases, outperform conventional algorithms by correctly resolving elemental identities and intricate structural motifs including hydrogen-bonding networks and moderate disorder. The success of *RefrActor* marks a conceptual shift from local, iterative refinement to a global, data-driven generative process. This transformation reframes structure determination as a problem of probabilistic inference rather than heuristic optimization, democratizing access for non-experts while accelerating discovery for experienced crystallographers. While challenges such as twinning and severe disorder define clear avenues for future work, the generative refinement paradigm sets the stage for fully autonomous and data-driven crystallographic analysis, paving the way toward an integrated pipeline for automated materials characterization.

## Author Information

### Corresponding Author.

* chli@nju.edu.cn; zhaoyue@nju.edu.cn; zuojl@nju.edu.cn.

### Author Contributions

The manuscript was written through contributions of all authors. All authors have given approval to the final version of the manuscript.

## Supplementary Information

All technical details, data curation criteria, and additional validation results associated with this work are thoroughly documented in the SI. Specifically, the SI provides: (1) the complete curation pipeline and filtering standards used to construct the GenRef-10k benchmark dataset (**Section S1**); (2) detailed pseudocode for key algorithms, including Reciprocal Space Voxelization and the Contrastive Pair Generation Strategy (**Section S2**); (3) full model architecture specifications, training procedures (**Section S3.10**), and the methods used for crystallographic symmetry representation and processing (**Section S3.11**); (4) comprehensive validation of the *ReciEncoder* model, including ablation studies of its hybrid architecture and multi-task training curriculum (**Sections S3.2–S3.3**); (5) an in-depth analysis of the learned embedding space and the model's generalization behavior (**Sections S3.4–S3.6**); (6) the crystallographic rationale and statistical characterization of the GenRef-10k benchmark and its "challenging subsets," together with detailed model performance across these subsets (**Sections S3.7–S3.9**); and (7) a case-study gallery of 30 full refinement examples, illustrating the model's end-to-end structure-solution capabilities across diverse chemical systems (**Supplementary Figures S25-S28**).

### Notes

The authors declare no competing financial interest.

### Acknowledgement

This work was supported by the National Natural Science Foundation of China (Grant No.

22425106 and 22494633), and the Natural Science Foundation of Jiangsu Province (BK20243010). We thank the Cambridge Crystallographic Data Centre (CCDC) for providing the CSD Python API (version 3.4.1) used in this work. We also thank Mrs. Sihan Yang from Nanjing Jinglide Technology Company Limited for her help in crystal structure determination.